\documentclass[journal]{IEEEtran}
\usepackage{cite}
\usepackage{threeparttable}
\usepackage{graphicx}
\usepackage{picinpar}
\usepackage{amssymb}
\usepackage[cmex10]{amsmath}
\usepackage{amsmath,amsfonts,amssymb}

\usepackage{subfigure}
\usepackage{algorithm}
\usepackage{algorithmic}
\usepackage{changebar}

\input epsf

\begin{document}

\title{Priori-Information Aided Iterative Hard Threshold: A Low-Complexity High-Accuracy
 Compressive Sensing Based Channel Estimation for TDS-OFDM}

\author{Zhen Gao, Chao Zhang,~\IEEEmembership{Member,~IEEE},
 Zhaocheng Wang,~\IEEEmembership{Senior Member,~IEEE},
 Sheng Chen,~\IEEEmembership{Fellow,~IEEE}%
\thanks{Z. Gao, C. Zhang, and Z. Wang are with Tsinghua National Laboratory for
 Information Science and Technology (TNList), Department of Electronic Engineering,
 Tsinghua University, Beijing 100084, China (E-mails: gao-z11@mails.tsinghua.edu.cn;
 z\_c@tsinghua.edu.cn; zcwang@mail.tsinghua.edu.cn).} %
\thanks{S. Chen is with Electronics and Computer Science, University of Southampton,
 Southampton SO17 1BJ, U.K. (E-mail: sqc@ecs.soton.ac.uk), and also with King
 Abdulaziz University, Jeddah 21589, Saudi Arabia.} %
\thanks{This work was supported by National Key Basic Research Program of
China (Grant No. 2013CB329203), National Nature Science Foundation
of China (Grant No. 61271266), National High Technology Research
and Development Program of China (Grant No. 2014AA01A704), and
Beijing Natural Science Foundation (Grant No. 4142027).}  %
\vspace*{-5mm}
}

\maketitle

\begin{abstract}
 This paper develops a low-complexity channel estimation (CE) scheme based on
 compressive sensing (CS) for time-domain synchronous (TDS) orthogonal
 frequency-division multiplexing (OFDM) to overcome the performance loss under
 doubly selective fading channels. Specifically, an overlap-add method of the
 time-domain training sequence is first proposed to obtain the coarse estimates
 of the channel length, path delays and path gains of the wireless channel, by
 exploiting the channel's temporal correlation to improve the robustness of the
 coarse CE under the severe fading channel with long delay spread. We then
 propose the priori-information aided (PA) iterative hard threshold (IHT)
 algorithm, which utilizes the priori information of the acquired coarse
 estimate for the wireless channel and therefore is capable of obtaining an
 accurate channel estimate of the doubly selective fading channel. Compared with
 the classical IHT algorithm whose convergence requires the $l_2$ norm of the
 measurement matrix being less than 1, the proposed PA-IHT algorithm exploits
 the priori information acquired to remove such a limitation as well as to reduce
 the number of required iterations. Compared with the existing CS based CE method
 for TDS-OFDM, the proposed PA-IHT algorithm significantly reduces the
 computational complexity of CE as well as enhances the CE accuracy. Simulation
 results demonstrate that, without sacrificing spectral efficiency and changing
 the current TDS-OFDM signal structure, the proposed scheme performs better than
 the existing CE schemes for TDS-OFDM in various scenarios, especially under
 severely doubly selective fading channels.
\end{abstract}

\begin{IEEEkeywords}
 Orthogonal frequency-division multiplexing (OFDM), digital terrestrial
 television broadcasting, time-domain synchronous OFDM, compressive sensing,
 channel estimation
\end{IEEEkeywords}

\IEEEpeerreviewmaketitle

\section{Introduction}\label{S1}

 Orthogonal frequency-division multiplexing (OFDM) technology has been widely
 applied in high-speed broadband wireless communication systems \cite{ofdm}.
 In the digital terrestrial television broadcasting (DTTB) field, both European
 second generation digital video broadcasting standard (DVB-T2) \cite{etsi2012}
 and Chinese digital terrestrial multimedia broadcasting standard (DTMB)
 \cite{GB2006} adopt OFDM as the key modulation technology. DVB-T2 uses cyclic
 prefix (CP) based OFDM, where a CP, serving as the guard interval, is inserted
 between successive OFDM data blocks to eliminate the inter-block-interference
 (IBI) caused by multipath channels \cite{IBI}. Unlike DVB-T2, DTMB uses
 time-domain synchronous (TDS) OFDM, which replaces the CP with a time-domain
 training sequence (TS). Compared with the classical CP based OFDM, TDS-OFDM has
 superior performance in terms of fast synchronization and channel estimation
 (CE), and it also achieves a higher spectral efficiency \cite{cuba,intera,DPN}.
 Owing to the good performance of TDS-OFDM, DTMB has been officially approved
 as an international DTTB standard, and has been successfully deployed in
 China and several other countries \cite{cuba}.

 However, because of the mutual interferences between the TS and the OFDM data
 block, an iterative interference cancellation has to be used in TDS-OFDM
 systems to decouple the TS and the OFDM data block for CE and frequency-domain
 demodulation \cite{intera}. This iterative interference cancellation will
 result in the performance degradation under doubly selective fading channels
 \cite{cuba}, whereby the perfect removal of the IBIs is difficult to realize.
 To solve this problem, several schemes have been proposed \cite{UW,tft,DPN}.
 Among these solutions, the dual pseudo-noise OFDM (DPN-OFDM) has attracted
 more attention owing to its simplicity and its ability to deliver an accurate
 channel estimate without imposing complex iterative interference cancellation
 \cite{DPN}, at the cost of sacrificing some spectral efficiency. Recently, a
 channel estimation method was proposed based on compressive sensing (CS) for
 the current DTMB system \cite{Dai}, whereby the small IBI-free region in the
 received TS is exploited to estimate the long multipath channel by exploiting
\begin{changebar}the greedy signal recovery algorithms for CS, such as the subspace
 pursuit (SP) algorithm \cite{SP} and the compressive sampling matching pursuit
 (CoSaMP) algorithm \cite{greedy}.\end{changebar} However, this scheme suffers from the
 drawback of high computational complexity owing to the matrix inversion operation
 required as well as the inevitable performance degradation under the adverse
 conditions of severe multipath channels with very long delay spread.

 Against this background, in this paper we develop a low-complexity high-accuracy
 CS based CE scheme for TDS-OFDM systems. Our contribution is twofold. Firstly,
 we derive the overlap-add method of the TS to obtain the coarse estimates of the
 channel length, path delays and path gains of the wireless channel, whereby the
 temporal correlation of the multipath delays and multipath gains among several
 consecutive TDS-OFDM symbols is exploited. More specifically, the TS tail part
 caused by the multipath channel is superposed on the preceding TS main part, and
 then this overlap-add result is circularly correlated with the local PN sequence
 to obtain some coarse channel state information (CSI). Moreover, the temporal
 correlation of the multipath delays and gains among several consecutive TDS-OFDM
 symbols is jointly exploited to improve the robustness and accuracy of the coarse
 CE. Assisted by the prior information of the wireless channel obtained by the
 overlap-add method, our proposed CS based CE method, referred to as the
 priori-information aided iterative hard threshold (PA-IHT), is capable of obtaining
 an accurate channel estimate while only imposing a low computational complexity.
 In particular, unlike the classical iterative hard threshold (IHT) algorithm
 \cite{IHT} whose convergence requires the $l_2$ norm of the measurement matrix
 being less than 1, our PA-IHT algorithm utilises the available priori information
 of the wireless channel to remove such a restriction as well as to reduce the
 number of required iterations. Also benefiting from the acquired priori information
 of the wireless channel, our PA-IHT algorithm significantly improves signal
 recovery accuracy as well as considerably reduces the computational complexity,
 in comparison to the the existing CS based CE method, such as the modified CoSaMP
 algorithm of \cite{Dai}. Moreover, unlike the modified CoSaMP algorithm which
 suffers from invertible CE performance degradation under severe multipath channels
 with very long delay spread, the PA-IHT algorithm remains robust and accurate under
 such adverse channel conditions.
\vspace*{-1mm}

 The rest of the paper is organised as follows. Section~\ref{S2} presents the
 TDS-OFDM system model and briefly reviews several existing CE schemes for
 TDS-OFDM. Section~\ref{S3} details our proposed PA-IHT based CE scheme for
 TDS-OFDM, while Section~\ref{S4} provides the simulation results to demonstrate
 the superior performance of our PA-IHT algorithm over the existing state-of-the-arts.
 Our conclusions are drawn in Section~\ref{S5}.
\vspace*{-1mm}

 Throughout our discussions, boldface capital and lower-case letters stand for
 matrices and column vectors, respectively. The exception is for frequency-domain
 vectors, whereby the discrete Fourier transform (DFT) of the time-domain vector
 $\mathbf{x}$ is denoted by $\boldsymbol{X}$. The operators $\ast$ and $\otimes$
 represent the linear convolution and circular correlation, respectively, while
 $\lfloor \cdot \rfloor$ denotes the integer floor operator and $\rm{abs}\{\bf{x}\}$
 is the vector whose elements are the absolute values of the corresponding
 elements of the vector $\mathbf{x}$. The transpose and conjugate transpose
 operations are denoted by $(\cdot )^{\rm T}$ and $(\cdot )^{\rm H}$, respectively,
 while $(\cdot )^{\dag}$ denotes the Moore-Penrose matrix inversion and the $l_p$
 norm operation is given by $\|\cdot\|_p$. The $r$-sparse vector of $\mathbf{x}$
 is denoted by $\mathbf{x}\rangle_r$ which is generated by retaining the $r$ largest
 elements of $\mathbf{x}$ and setting the rest of the elements to zero. The support
 of the vector $\mathbf{x}$ is denoted by ${\rm supt} \{\mathbf{x}\}$, and
 $\left. \mathbf{x} \right|_{\Gamma}$ denotes the entries of $\mathbf{x}$ defined
 in the set $\Gamma$, while $\left.\mathbf{\Phi}\right|_{\Gamma}$ denotes the sub-matrix
 whose columns comprise the columns of $\mathbf{\Phi}$ that are defined in the set
 $\Gamma$. \begin{changebar} Additionally, $| \cdot |$ denotes the absolute value,
 and $| \cdot |_{\rm c}$ is the cardinality of a set.\end{changebar} Finally, $\delta (\cdot )$
 represents the unit impulse function.

\section{System Model}\label{S2}

 In the time domain, TDS-OFDM signals are grouped in symbols, and each TDS-OFDM symbol
 consists of a TS, which is a known pseudo-noise (PN) sequence $\mathbf{c}=\big[c_0 ~ c_1
 \cdots c_{M-1}\big]^{\rm T}$ of length $M$, and the following OFDM data block of
 length $N$, which can be expressed as $\mathbf{x}_i=\big[x_{i,0} ~ x_{i,1} \cdots
 x_{i,N-1}\big]^{\rm T}$ with $i$ denoting the TDS-OFDM symbol index. Hence, the
 $i$th TDS-OFDM symbol is represented by $\mathbf{s}_i=\big[\mathbf{c}^{\rm T} ~
 \mathbf{x}_i^{\rm T}\big]{\rm ^T}=\big[\mathbf{c}^{\rm T} ~ \big(\mathbf{F}_N^{\rm H}
 \boldsymbol{X}_i\big)^{\rm T}\big]^{\rm T}$, where $\mathbf{F}_N$ is the DFT matrix
 of size $N\times N$ and $\boldsymbol{X}_i=\big[ X_{i,0} ~ X_{i,1} \cdots
 X_{i,N-1}\big]^{\rm T}$ is the frequency-domain $i$th OFDM data block.

 At the receiver, the received $i$th OFDM symbol can be written as $\mathbf{r}_i=
 \mathbf{s}_i\ast \mathbf{h}_i + \mathbf{n}_i$, where $\mathbf{n}_i$ is the channel
 additive white Gaussian noise (AWGN) vector having a zero mean, while $\mathbf{h}_i
 =\big[h_{i,0} ~ h_{i,1}\cdots h_{i,L-1}\big]^{\rm T}$ is the time-varying channel
 impulse response (CIR) of length $L$ which can be considered to be quasi-static
 during the time period of the $i$th TDS-OFDM symbol. Since the wireless channel is
 sparse in nature \cite{nature}, its CIR comprises only $P$ resolvable propagation
 paths, where $P \ll L$. In other words, only $P$ coefficients of $\mathbf{h}_i$
 are non zero, and therefore the coefficients of the CIR can be expressed by the
 following model \cite{sCIR1,sCIR2}
\begin{align}\label{eq1}
 h_{i,l} =& \sum\limits_{p=0}^{P-1} \alpha_{i,p}\delta\big(l-\tau_{i,p}\big) , \,
 0\le l \le L-1 ,
\end{align}
 where $\alpha_{i,p}$ is the $p$th path gain and $\tau_{i,p}$ is the $p$th path delay.
 Obviously, we have
\begin{align}\label{eq2}
 h_{i,l} =& \left\{ \begin{array}{cl} \alpha_{i,p}, & \mbox{if }l = \tau_{i,p} , \\
  0 , & \mbox{otherwise} .   \end{array} \right.
\end{align}

\begin{changebar}
\begin{figure}[b!]
\vspace*{-2mm}
\begin{center}
 \includegraphics[width=\columnwidth, keepaspectratio]{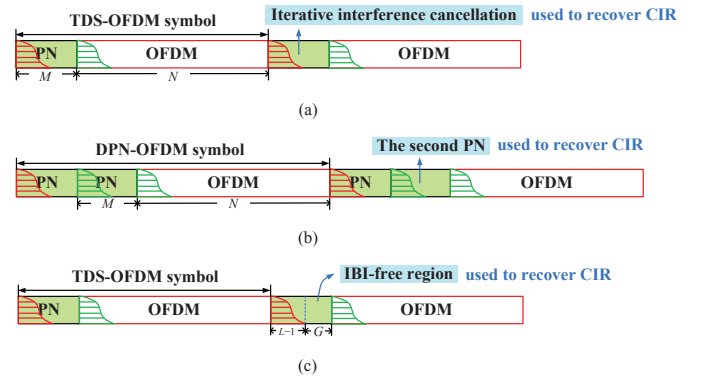}
\end{center}
\vspace*{-4mm}
\caption{Existing channel estimation schemes for TDS-OFDM: (a)~the scheme based
 on iterative interference cancellation, (b)~the scheme based on dual PN sequences,
 and (c)~the scheme based on compressive sensing.}
\label{fig:tds_ofdm} % Fig 1
\end{figure}
\end{changebar}

\begin{changebar}
\begin{figure*}[tp!]
\vspace*{-1mm}
\begin{center}
 \includegraphics[width=\linewidth, keepaspectratio]{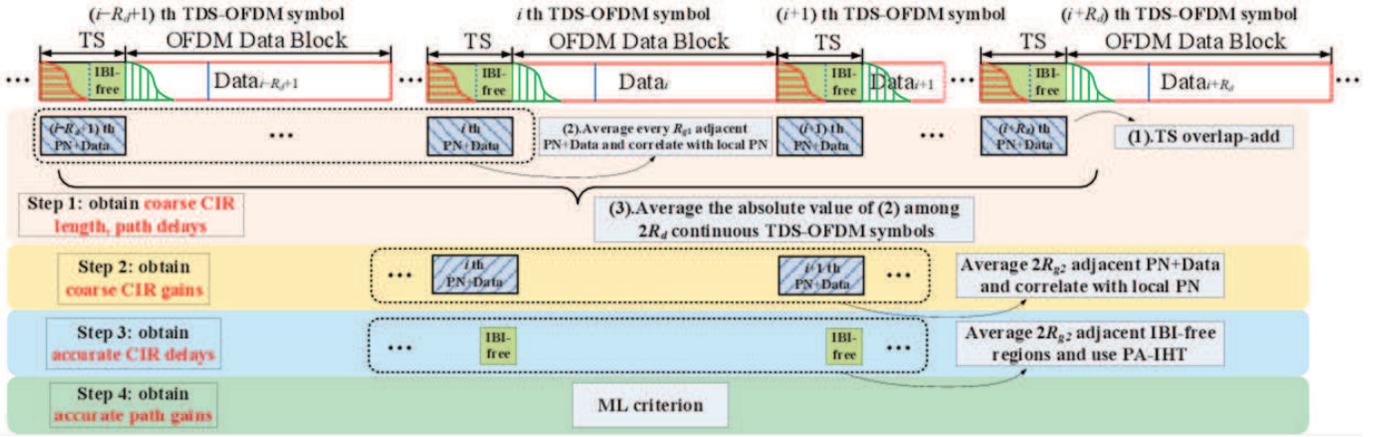}
\end{center}
\vspace*{-3mm}
\caption{The proposed PA-IHT based channel estimation which consists of four steps.
 The first two steps use the proposed overlap-add method of the TS, whereby the
 temporal correlation of the wireless channel is exploited to obtain some priori
 information of the wireless channel. In the rest two steps, the proposed PA-IHT
 algorithm and the ML criterion are used to obtain an accurate channel estimation.}
 \label{fig:duozhenhebing} % Fig2
\vspace*{-3mm}
\end{figure*}
\end{changebar}

 For CE and data demodulation in TDS-OFDM, the existing schemes may not be able
 to achieve satisfactory performance, especially under severe multipath channels.
 Fig.~\ref{fig:tds_ofdm} illustrates several existing CE schemes for TDS-OFDM
 systems. As shown in Fig.~\ref{fig:tds_ofdm}~(a), the conventional CE scheme
 for TDS-OFDM using single PN sequence has the advantage of maintaining a high
 spectral efficiency. However, this scheme suffers from the mutual interferences
 between the PN sequence and the OFDM data block. Therefore, an iterative
 interference cancellation has to be adopted to decouple the mutual interferences
 \cite{intera}, which reduces the accuracy of CE in the doubly selective fading
 channel \cite{cuba}. In the DPN-OFDM scheme, as shown in Fig.~\ref{fig:tds_ofdm}~(b),
 an extra PN sequence is inserted to prevent the second PN sequence from being
 contaminated by the preceding OFDM data block. In this way, the DPN-OFDM scheme
 removes the complex iterative interference cancellation and improves the CE
 performance, but at the cost of reducing the spectral efficiency.

 The PN sequence length in a TDS-OFDM system is designed to be longer than the maximum
 CIR length in order to ensure the reliable system performance. Considering the
 wireless scenarios that the actual CIR length $L$ is less or even much less than the
 length of the guard interval $M$, there is an IBI-free region of the small size
 $G=M-L+1$ at the end of the received PN sequence, as illustrated in
 Fig.~\ref{fig:tds_ofdm}~(c). In this IBI-free region, the received signal $\mathbf{y}=
 \big[y_{L-1} ~ y_L \cdots y_{M-1}\big]^{\rm T}$ can be represented by
\begin{align}\label{equ:cs_euq} % eq3
 \mathbf{y} =& \mathbf{\Phi}\mathbf{h} + \mathbf{n}^{'} ,
\end{align}
 where $\mathbf{n}^{'}$ is the related channel AWGN vector, and
\begin{align}\label{equ:phi2} % eq4
 \mathbf{\Phi} =& \left[ \begin{array}{*{20}{c}}
  c_{L-1} & c_{L-2} & \cdots & c_0 \\
  c_L & c_{L-1} & \cdots & c_1 \\
  \vdots & \vdots & \vdots & \vdots \\
  c_{M-1} & c_{M-2} & \cdots & c_{M-L} \\
 \end{array} \right]_{G\times L}
\end{align}
 is a Toeplitz matrix of size $G \times L$ which is completely determined by the TS.

 Generally speaking, the IBI-free region, as shown in Fig.~\ref{fig:tds_ofdm}~(c),
 is usually small. Thus it is difficult to obtain a unique solution to the unknown
 channel $\mathbf{h}$ in (\ref{equ:cs_euq}), since the observation dimension $G$
 is usually smaller than the CIR dimension $L$. Fortunately, the compressive
 sensing theory \cite{greedy} has proved that the high dimension signal can be
 accurately reconstructed by the low dimensional uncorrelated observations if the
 target signal is sparse or approximately sparse. A wireless channel is sparse in
 nature \cite{nature} and the actual number of the resolvable paths $P$ usually
 satisfies $P\ll L$. Therefore, even though the CIR dimension $L$ is larger or
 even much larger than the observation dimension $G$, we may have $P \le G$. In
 \cite{Dai}, a modified CoSaMP algorithm is proposed to solve the under-determined
 equation (\ref{equ:cs_euq}). This scheme inherits the advantage of high spectral
 efficiency without changing the current TDS-OFDM signal structure. Furthermore,
 it improves the CE performance without the need of iterative interference
 cancellation. However, this scheme has a high computational complexity due to the
 required matrix inversion operations in the CS algorithm and moreover its
 performance degrades over severe multipath channels with long delay spread, where
 the observation dimension $G$ may become too small.

\section{PA-IHT Based Channel Estimation}\label{S3}

 In this section, we present the CE method based on the proposed PA-IHT algorithm for
 TDS-OFDM systems and also provide the complexity analysis for this CE scheme.

\subsection{The Proposed PA-IHT Based Channel Estimation Method}\label{S3.1}

 The proposed CE method consists of four steps, as shown in Fig.~\ref{fig:duozhenhebing}.
 To be more specific, in the first step, a coarse CIR length and path delays are estimated,
 while in the second step, coarse channel gains are obtained. With the aid of the  coarse
 information of the wireless channel obtained in the first two steps, the proposed PA-IHT
 algorithm estimates the accurate path delays in the third step, and finally at the fourth
 step, the accurate path gains are obtained based on a maximum likelihood (ML) criterion.

\begin{figure}[hp!]
\begin{center}
 \includegraphics[width=\columnwidth, keepaspectratio]{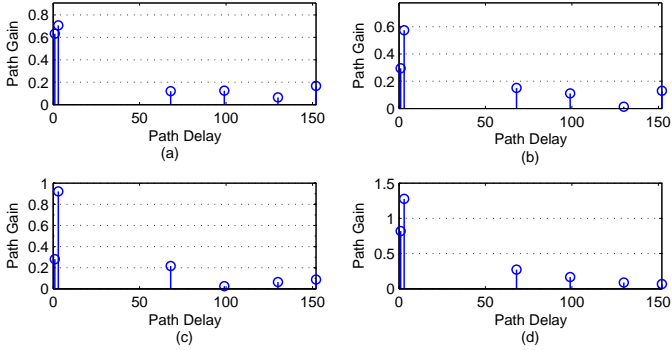}
\end{center}
\vspace*{-4mm}
\caption{The CIRs of four adjacent TDS-OFDM symbols over the ITU-VB channel with
 120km/h receiver velocity. The DTTB carrier frequency $f_c=634\rm{MHz}$ and the
 symbol rate $1/T_s=7.56\rm{MHz}$.}
 \label{fig:dongtaiduojing} % Fig 3
\end{figure}

 In the first two steps, we exploit the temporal correlation of the wireless channel to
 estimate the coarse CIR length, path delays, and path gains. For time-varying  channels,
 the path delays usually vary more slowly than the path gains \cite{Channel_corr}. Even
 for mobile scenarios, although the path gains will change over adjacent TDS-OFDM symbols,
 the path delays may remain relatively unchanged. \begin{changebar}This is because the duration
 $T_{\text{delay}}$ for the delay of a particular path to change by one tap is inversely
 proportional to the signal bandwidth $f_s$, while the coherence time of the path gains
 $T_{\text{gain}}$ is inversely proportional to the carrier frequency $f_c$
 \cite{Channel_corr,wide}. Since $f_s \ll f_c$ for all practical wireless systems,
 path delays change much slower than path gains, i.e. $T_{\text{gain}}\ll T_{\text{delay}}$.\end{changebar}
 Fig.~\ref{fig:dongtaiduojing} depicts the CIRs of four adjacent TDS-OFDM symbols over
 International Telecommunications Union Vehicular B (ITU-VB) channel \cite{Chan} with
 120km/h receiver velocity. From Fig.~\ref{fig:dongtaiduojing}, we observe that although
 the path gains are different in adjacent TDS-OFDM symbols, the path delays remain nearly
 invariant. The characteristics of the time-varying channel over several TDS-OFDM symbols
 can therefore be exploited to assist the coarse CE.

\begin{changebar}Let us now elaborate this formally. Firstly, for a time-varying channel, the path
 delays of the channel in the time interval length of $T_{\text{delay}}$, or the duration of
 $\left\lfloor \frac{T_{\text{delay}}}{T_s(M + N)}\right\rfloor$ TDS-OFDM symbols, can be
 considered to remain nearly unchanged \cite{Channel_corr,wide}. In other words, the CIR
 can be considered to share the same sparsity pattern during $2{R_d}-1$ TDS-OFDM symbols
 \cite{MMV}, where $R_d=\left\lfloor \frac{T_{\text{delay}}}{2T_s(M + N)} \right\rfloor$,
 or in the time duration of $T_{\text{delay}}$ the temporally common sparsity of the
 wireless channel is guaranteed \cite{wide}. In practice, $R_d$ can be very large since
 $T_{{\rm{delay}}}\propto \frac{{T_s}c}{v}=\frac{c}{{f_s}v}$, where $c$ is the velocity
 of light and $v$ is the speed of mobile receiver, while $T_s$ is the data symbol duration
 which may therefore be regarded as the resolution of delays. Hence we may obtain
 approximately ${R_d}\approx \left\lfloor\frac{c}{2v(M + N){f_sT_s}}\right\rfloor$. Taking
 the example of $v=100~\rm{m/s}$ and $M+N=256+4096$ for instance, $R_d$ can be considered
 to be unchanged during a superframe in the DTMB system, where a DTMB superframe consists
 of multiple TDS-OFDM symbols.

 Secondly, over the coherence time of the path gains, $T_{\text{gain}}$, the channel gains
 can be expressed as $\left| \alpha_{i,p} \right|\exp \big(\phi_0 + 2\pi f_d t\big)$,
 where $\phi_0$ is the initial phase, $t$ denotes the time, and $f_d$ is the Doppler
 frequency offset which can be estimated at the receiver \cite{dopller1}.\end{changebar}
 Hence, the phase variation of a complex path gain is less than $\pi$ within the time
 interval of $\frac{1}{2f_d}$ length, or the duration of $R_{g1}=\left\lfloor
 \frac{1}{2 f_d T_s(M + N)} \right\rfloor$ TDS-OFDM symbols. \begin{changebar}In other words,
 the path delays of the channel are high correlated during $R_{g1}=\left\lfloor
 \frac{1}{2 f_d T_s(M + N)} \right\rfloor$ TDS-OFDM symbols.\end{changebar} By averaging the CIR
 estimate over $R_{g1}$ adjacent TDS-OFDM symbols, therefore, we can reduce the effects
 of the channel AWGN and improve the accuracy and reliability of the path delay estimation.
\begin{changebar}Clearly, for this averaging to be effective, we must have $R_{g1} > 1$.\end{changebar}

\begin{changebar}It is well known that $f_d\propto \frac{{f_c}v}{c}$.  Hence we may obtain
 approximately $R_{g1}\approx \left\lfloor\frac{c}{2v(M + N){{f_c}{T_s}}}\right\rfloor$.
 Since $f_s \ll f_c$, clearly we have $2R_d - 1 > R_{g1} > 1$. In practice, the receiver
 can adaptively choose appropriate values for the parameters $R_{d}$ and $R_{g1}$ based on
 the channel status and the estimated $f_d$ of the time-varying channel.\end{changebar}

 Furthermore, in order to achieve the reliable coarse channel estimation in the case of
 instantaneous deep channel fading occurring during $R_{g1}$ adjacent TDS-OFDM symbols, the
 CIR estimations over the $2{R_d}-1$ adjacent TDS-OFDM symbols are jointly exploited to
 further improve the coarse estimation of the channel length and path delays.

\begin{changebar}
 Lastly, let us define the wireless channel being quasi-static during the duration of
 $2R_{g2}-1$ TDS-OFDM symbols. In general, we must assume that the path delays and
 path gains of the wireless channel remain unchanged at least during one TDS-OFDM symbol.
 Therefore, for time-varying channels, we can choose $R_{g2}=1$.
\end{changebar}

\begin{changebar}
 For static channels in particular where $f_d=0$, both the path delays and path gains
 are time-invariant. We can simple choose a desired value of $R_{g1} > 1$ for averaging,
 and further set $2R_d - 1 = 2R_{g2} - 1 = R_{g1}$.\end{changebar}
 We now detail our proposed PA-IHT algorithm.

\begin{changebar}
\begin{figure}[b!]
\vspace*{-3mm}
\begin{center}
 \includegraphics[width=\columnwidth, keepaspectratio]{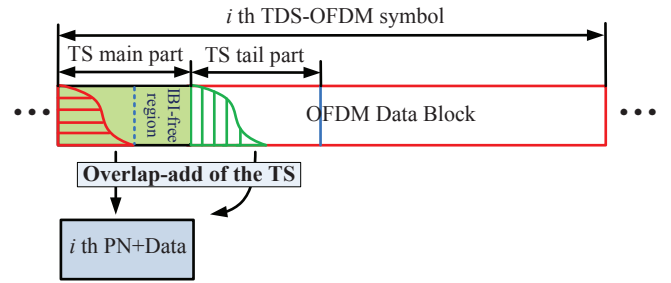}
\end{center}
\vspace*{-3mm}
\caption{Illustration for overlap-add of the TS in the $i$th TDS-OFDM symbol. Note that
 in \emph{Step 1}, the length of the TS tail part is $M$, while in \emph{Step 2}, the
 length of the TS tail part is the estimated CIR length.}
\label{fig:overlap} % Fig 4
\end{figure}
\end{changebar}

\subsubsection{Step 1. Acquisition of Coarse Channel Length and Path Delays}

 We propose the overlap-add method of the TS, which jointly utilizes the
 received TSs from the ($i-R_d+1$)th to ($i+R_d$)th TDS-OFDM symbols to exploit
 the temporal correlation of the wireless channel. The proposed overlap-add
 method of the TS is illustrated in Fig.~\ref{fig:overlap}, and its operation
 is represented by
\begin{align}\label{equ:recon} % eq5
 \mathbf{r}_k =& \mathbf{r}_{k,{\rm main}} + \mathbf{r}_{k,{\rm tail}} , \,
 i - R_d+1 \le k \le i + R_d ,
\end{align}
 in which the TS main part $\mathbf{r}_{k,{\rm main}}$ and the TS tail part
 $\mathbf{r}_{k,{\rm tail}}$ can be expressed respectively by
\begin{align}
 \mathbf{r}_{k,{\rm main}} =& \mathbf{\Psi}_k \mathbf{h}_k + \mathbf{n}_{k,{\rm main}} ,
 \, i - R_d +1\le k \le i + R_d , \label{equ:main} \\  % eq6
 \mathbf{r}_{k,{\rm tail}} =& \mathbf{\Theta}_k \mathbf{h}_k + \mathbf{n}_{k,{\rm tail}} ,
 \, i - R_d +1\le k \le i + R_d , \label{equ:tail} % eq7
\end{align}
 where $\mathbf{n}_{k,{\rm main}}$ and $\mathbf{n}_{k,{\rm tail}}$ are the corresponding
 AWGN vectors, while
\begin{align}\label{eq8}
 \mathbf{\Psi}_k \! =& \! \left[ \!\! \begin{array}{ccccc}
  c_0 \!\! & \!\! x_{k-1,N-1} \!\! & \!\! x_{k-1,N-2} \!\! & \!\! \cdots \!\! & \!\! x_{k-1,N-L+1} \\
  c_1 \!\! & \!\! c_0 \!\! &  \!\!x_{k-1,N-1} \!\! & \!\! \cdots \!\! & \!\! x_{k-1,N-L+2} \\
  \vdots \!\! & \!\! \vdots \!\! & \!\! \vdots \!\! & \!\! \ddots \!\! & \!\! \vdots \\
  c_{L-1} \!\! & \!\! c_{L-2} \!\! & \!\! c_{L-3} \!\! & \!\! \cdots \!\! & \!\! c_0 \\
  \vdots \!\! & \!\! \vdots \!\! & \!\! \vdots \!\! & \!\! \ddots \!\! & \!\! \vdots \\
  c_{M-1} \!\! & \!\! c_{M-2} \!\! & \!\! c_{M-3} \!\! & \!\! \cdots \!\! & \!\! c_{M-L}
\end{array} \!\! \right]_{M \times L} \! , \!
\end{align}
\begin{align}\label{eq9}
 \mathbf{\Theta}_k \! =& \! \left[ \begin{array}{ccccc}
 x_{k,0} & c_{M-1} & c_{M-2} & \cdots & c_{M-L+1} \\
 x_{k,1} & x_{k,0} & c_{M-1} & \cdots & c_{M-L+2} \\
 \vdots & \vdots & \vdots & \ddots & \vdots \\
 x_{k,L-1} & x_{k,L-2} & x_{k,L-3} & \cdots & x_{k,0} \\
 \vdots & \vdots & \vdots & \ddots & \vdots \\
 x_{k,M-1} & x_{k,M-2} & x_{k,M-3} & \cdots & x_{k,M-L}
\end{array} \right]_{M \times L} \! . \!
\end{align}

 Subsequently, the overlap-add results of the TS of $R_{g1}$ adjacent TDS-OFDM
 symbols are averaged, and then circularly correlated with the known TS,
 whereby the good auto-correlation and circular cross-correlation property of
 the TS is exploited. Specifically, we have \begin{changebar}
\begin{align}\label{equ:circ_conv} % eq10
 \widetilde{\mathbf{h}}_q =& \frac{1}{MR_{g1}} \left( \mathbf{c} \otimes
 \sum\limits_{k=q}^{q+R_{g1}-1} \mathbf{r}_k \right) \nonumber  = \frac{1}{R_{g1}}
 \sum\limits_{k=q}^{q+R_{g1}-1} \mathbf{h}_k + \mathbf{v}_k , \, \nonumber \\ &
 i - R_d + 1 \le q \le i + R_d - R_{g1} ,
\end{align}
 where $\mathbf{v}_k$ represents the circular correlation of the interference
 plus the channel AWGN with the PN sequence averaging over the TS of $R_{g1}$
 adjacent TDS-OFDM symbols. Note that by averaging over $R_{g1}$ adjacent TDS-OFDM
 symbols, the effect of the AWGN is significantly reduced.\end{changebar} Consequently, the
 coarse CE $\bar{\mathbf{h}}$ is given by\begin{changebar}
\begin{align}\label{equ:coarse_chan} % eq11
 \bar{\mathbf{h}} =& \frac{1}{2R_d - R_{g1}}\sum\limits_{q=i-R_d+1}^{i+R_d-R_{g1}}
 \rm{abs}\big\{ \widetilde{\mathbf{h}}_q\big\}  .
\end{align}
 For time-varying channels, $2R_d - R_{g1}( > 1)$ estimates $\mathbf{h}_q$ are
 utilized to obtain the coarse CE $\bar{\mathbf{h}}$, and this reduces the
 effect of instantaneous deep channel fading occurring during the duration of a
 particular TDS-OFDM symbol. For static channels, $2R_d - R_{g1} = 1$, and
 only single estimate $\mathbf{h}_q$ is used to obtain the coarse CE
 $\bar{\mathbf{h}}$, since the channel delays and gains are constant over
 adjacent TDS-OFDM symbols.\end{changebar} Finally, only the propagation path
 delays of the most significant taps
\begin{align}\label{eq12}
 D_0=\big\{ \tau_1 : \big| \bar{h}_{\tau_1}\big| \ge E_{\rm th}\big\}_{\tau_1=0}^{L-1}
\end{align}
 are retained, and the retained $\big\{\bar{h}_{\tau_1}\big\}_{\tau_1=0}^{L-1}$ form
 the resulting coarse estimate $\bar{\mathbf{h}}$, where $E_{\rm th}$ is the power
 threshold which can be determined according to \cite{th}. In this way, the channel
 length can be estimated from the coarse CE according to
\begin{align}\label{eq13}
 \widehat{L} =& \max_{\tau_1\in D_0} \tau_1 + a ,
\end{align}
\begin{changebar}where $a$ is a variable parameter used to define the IBI-free region comprising
 the last $G$ samples of the received TS, which can be determined according to \cite{wide}.

 With an initial channel sparsity level given by $S_0=\left| D_0 \right|_{\rm c}$,\end{changebar} the
 channel sparsity level is then determined according to $S=S_0+b$, \begin{changebar}where $b$ is a
 positive number used to combat the interference effect, since some low power paths
 may be treated as noise, and the value of $b$ can be calculated according to \cite{wide}.\end{changebar}
 Effectively, $S$ is a coarse estimate of the number of resolvable propagation paths $P$.

\subsubsection{Step 2. Acquisition of Coarse Channel Path Gains}

 The received TSs from ($i-R_{g2}+1$)th to ($i+R_{g2}$)th TDS-OFDM symbols are then
 used to provide the coarse estimate of the channel path gains according to
\begin{align}\label{equ:h_gains} % eq14
 \bar{\mathbf{h}}^{'} =& \mathbf{c} \otimes \frac{1}{2R_{g2}M} \sum\limits_{k=i-R_{g2}+1}^{i+R_{g2}}
 \left(\mathbf{r}_{k,{\rm main}} + \mathbf{r}^{'}_{k,{\rm tail}}\right) ,
\end{align}
 where $\mathbf{r}^{'}_{k,{\rm tail}}$ is the vector whose first $\widehat{L}$
 elements are the first $\widehat{L}$ of $\mathbf{r}_{k,{\rm tail}}$, while its
 rest elements are all zeros.

 The coarse estimates of the channel length, the channel path delays and path gains
 acquired in \emph{Steps 1} and \emph{2} provide the priori information of the wireless
 channel to assist the accurate CE using the PA-IHT algorithm in the following two
 steps.

\subsubsection{Step 3. Acquisition of Accurate Path Delay Estimate}

 We now present the proposed PA-IHT algorithm which utilizes the priori information
 from the coarse CE to improve the signal recovery accuracy and to reduce the
 computational complexity. Define the measurement vector as
\begin{align}\label{eq15}
 \bar{\mathbf{y}} =& \frac{1}{2R_{g2}} \sum\limits_{k=i-R_{g2}+1}^{i+R_{g2}}
 \mathbf{y}_k ,
\end{align}
 where $\mathbf{y}_k$ is the received signal vector in the IBI-free region of the
 $k$th TDS-OFDM symbol as given in (\ref{equ:cs_euq}) but its size is
 $\widehat{G}=M-\widehat{L}+1$. The corresponding Toeplitz matrix $\mathbf{\Phi}$
 of size $\widehat{G}\times \widehat{L}$ can be formed according to (\ref{equ:phi2}).
 The pseudocode of the proposed PA-IHT algorithm is summarized in Algorithm 1. The
 final estimated channel path delays are $D=\big\{ \tau_2: \big|\widehat{h}_{\tau_2}\big| > 0
 \big\}_{\tau_2=0}^{L-1}$, with $\big\{\widehat{h}_{\tau_2}\big\}_{\tau_2= 0}^{L-1}$
 being the elements of $\widehat{\mathbf{h}}$.

\begin{algorithm}[htb]
\renewcommand{\algorithmicrequire}{\textbf{Input:}}
\renewcommand\algorithmicensure {\textbf{Output:} }
\caption{ Priori-Information Aided Iterative Hard Threshold (PA-IHT).}
\label{alg:Framwork} % Alg 1
\begin{algorithmic}[1]
\REQUIRE
~~1) Initial path delay set $D_0$, coarse channel estimation
     $\bar{\mathbf{h}}^{'}$, channel sparsity level $S$; \\
     ~~~~~~2) Noisy measurements $\bar{\mathbf{y}}$, observation matrix $\mathbf{\Phi}$.
\ENSURE
 S-sparse estimation $\widehat{\mathbf{h}}$. \\

\STATE $\left. \mathbf{x}^0 \right|_{D_0} \leftarrow \left. \bar{\mathbf{h}}^{'} \right|_{D_0}$;

\STATE $u_{\rm current} = \left\| \bar{\mathbf{y}} - \mathbf{\Phi}\mathbf{x}^0 \right\|_2$;

\STATE $u_{\rm previous} = 0$;

\WHILE{$u_{\rm previous} \le u_{\rm current}$,}

\STATE $k \leftarrow k + 1$;

\STATE $\mathbf{z} = \mathbf{x}^{k-1} + \mathbf{\Phi}^{\rm H}\big(\bar{\mathbf{y}} -
 \mathbf{\Phi}\mathbf{x}^{k-1}\big)$;

\STATE $\Gamma = \sup \left\{\rm{abs}\{\mathbf{z}\}\rangle_S\right\}$;

\STATE $\mathbf{x}^k \leftarrow \mathbf{x}^{k-1}$;

\STATE $\left. \mathbf{x}^k \right|_\Gamma \leftarrow \left. \bar{\mathbf{h}}^{'} \right|_\Gamma$;

\STATE $\mathbf{x}^k \leftarrow {\mathbf{x}^k}\rangle_S$;

\STATE $u_{\rm previous} = u_{\rm current}$;

\STATE $u_{\rm current} = \left\| \bar{\mathbf{y}} - \mathbf{\Phi}\mathbf{x}^k \right\|_2$;

\ENDWHILE

\STATE $\widehat{\mathbf{h}} \leftarrow \mathbf{x}^{k-1}$. \\
\end{algorithmic}
\end{algorithm}

 In contrast to the classical IHT algorithm or other CS based algorithms, the
 proposed algorithm has several attractive features. Firstly, the PA-IHT
 algorithm exploits the available priori information of the coarse path delays
 and gains (or equivalently the locations and values of the partial large
 components in the target signal) as the initial condition, and this
 significantly enhances the signal recovery accuracy and reduces the number of
 required iterations. Secondly, unlike the modified CoSaMP algorithm \cite{Dai},
 the sizes of the IBI-free region and the measurement matrix are adaptively
 determined by the coarse channel length estimate $\widehat{L}$. Thirdly, the
 coarse path gains serve as the nonzero element values of the target signal in
 every iteration. By contrast, in order to obtain these values, the modified
 CoSaMP algorithm has to apply the least squares estimation with high-complexity
 matrix inversion operation while the classical IHT algorithm uses the correlated
 results of the measurement matrix and the residual error \cite{IHT}, whose
 convergence condition requires that $\left\|\mathbf{\Phi}\right\|_2 < 1$.

\subsubsection{Step 4. Accurate Path Gain Estimation Based on ML}

 The final channel estimate is obtained as the ML estimate \cite{Dai}
\begin{align}\label{equ:ML_} % eq16
 \left. \widehat{\mathbf{h}}^{'} \right|_D
 =& \big(\left.\mathbf{\Phi}\right|_D\big)^{\dag} \bar{\mathbf{y}}
 = \left( \big(\left.\mathbf{\Phi}\right|_D\big)^{\rm H} \left.\mathbf{\Phi}\right|_D
 \right)^{-1} \big(\left.\mathbf{\Phi}\right|_D\big)^{\rm H} \bar{\mathbf{y}} ,
\end{align}
 where $\widehat{\mathbf{h}}^{'}$ is a vector of length $M$, whose elements
 outside the set $D$ are zeros. We also give the Cramer-Rao lower bound (CRLB)
 of the proposed CE method \cite{Dai}
\begin{align}\label{equ:CRLB} % eq17
 {\rm Var}_{\rm CRLB} =& E\left\{ \left\| \bar{\mathbf{h}}^{'} - \mathbf{h} \right\|_2 \right\}
  = \frac{S}{2R_{g2}G \rho} ,
\end{align}
 where $\rho$ is the signal to noise ratio (SNR).

\subsection{Convergence Properties}\label{S3.2}

 In contrast to the conventional greedy algorithms for CS \cite{greedy}
 which are generally used to solve the problem of
\[ \min\limits_{\mathbf{x}} \left\{ \mathbf{x} \text{ is an }S\text{-sparse vector}:
 \left\| \mathbf{y} -\mathbf{\Phi}\mathbf{x} \right\|_2 \right\} , \]
 the proposed PA-IHT algorithm in \emph{Step 3} is used for the support detection.
 This support detection can be written as
\begin{align}\label{equ:Supp_detect} % eq18
 \min\limits_D \left\{ D: \left\| \mathbf{y} - \left. \mathbf{\Phi} \right|_D \left.
 \bar{\mathbf{h}}^{'} \right|_D \right\|_2 \right\} ,
\end{align}
 where $D$ is an $S$-dimension set, whose elements are in ascending order and
 they are specified by the indexes of the elements in $\bar{\mathbf{h}}^{'}$.
 Obviously, $D$ is uniquely determined when $\mathbf{y}$, $\mathbf{\Phi}$ and
 $\bar{\mathbf{h}}^{'}$ are given. The proposed algorithm solves the problem of
 (\ref{equ:Supp_detect}) in a greedy manner. When the iterative procedure meets
 the stopping criterion, the algorithm converges to at least a locally optimal
 solution.

 Moreover, from (\ref{equ:h_gains}), it is clear that $\bar{\mathbf{h}}^{'}$ is
 an unbiased estimate of the channel, because the data part mixed in the
 overlap-add result of the TS can be regarded as a noise with zero mean.
 Thus the estimate of the coarse channel delays in \emph{Step 3} contains the
 true channel path delays with high probability \cite{th}. By exploiting the
 priori information of the coarse channel path delays, the number of required
 iterations can be reduced, and the detected support tends to be a globally
 optimal solution.

\subsection{Computational Complexity}\label{S3.3}

 \emph{Steps 1} and \emph{2} implement the $M$-point circular correlation using fast Fourier
 transform (FFT), whose complexity is in the order of $\mathcal{O}\big( (M\log_2 M)/2\big)$.
 In \emph{Step 3}, owing to the priori information of the acquired coarse channel gains, our
 algorithm avoids the matrix inversion operation. In \emph{Step 4}, the ML estimate requires
 the  matrix inversion operation with the complexity of $\mathcal{O}\big(GS^2+S^3\big)$.
 Obviously, the main computational burden comes from \emph{Step 4}. and the complexity of our
 proposed algorithm is $C_{\rm PA-IHT}=\mathcal{O}\big(GS^2+S^3\big)$.

 The conventional CoSaMP algorithm and the modified CoSaMP algorithm can be shown
 to have the computational complexity of $C_{\rm CoSaMP}=\mathcal{O}\big(4GS^3+8S^4\big)$
 and $C_{\rm mCoSaMP}=\mathcal{O}\big((S-S_0)(4GS^2+8S^3)\big)$, respectively \cite{Dai}.
 The main computational burden of those two algorithms comes from the matrix inversion
 operation required to obtain the sparsity information or nonzero element values in the
 target signal. By contrast, our algorithm acquires this information at the cost of very
 low complexity of $\mathcal{O}\big( (M\log_2 M)/2\big)$.

 Considering the typical case of the ITU-VB channel \cite{Chan} where we have $S=6$,
 $G=104$ and $S_0=3$. The computational complexity of the three schemes are given
 respectively by $C_{\rm PA-IHT}=\mathcal{O}(3960)$, $C_{\rm CoSaMP}=\mathcal{O}(100224)$
 and $C_{\rm mCoSaMP}=\mathcal{O}(50112)$. We then have $C_{\rm PA-IHT}\big/C_{\rm CoSaMP}
 \approx 4\%$ and  $C_{\rm PA-IHT}\big/C_{\rm mCoSaMP}\approx 8\%$.

\begin{figure}[bp!]
\vspace*{-2mm}
\begin{center}
\includegraphics[width=0.9\columnwidth, keepaspectratio]{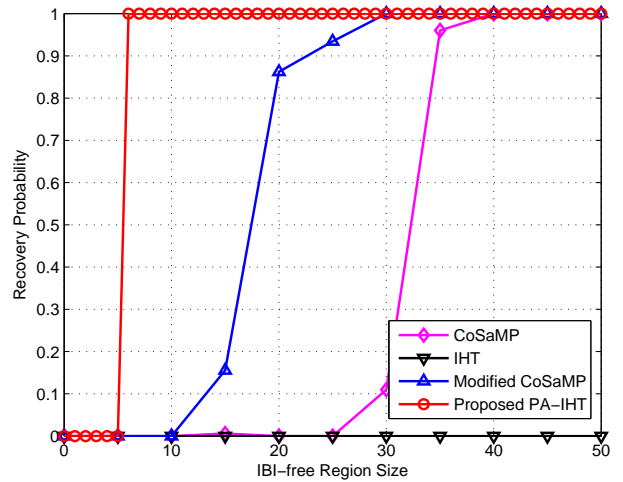}
\end{center}
\vspace*{-4mm}
\caption{Target signal recovery probabilities versus IBI-free region size attained
 by the four schemes for the static ITU-VB channel given $\mbox{SNR}=20$\,dB.}
\label{fig:add} % Fig 5
\end{figure}

\section{Simulation Results}\label{S4}

 A simulation study was carried out to compare the performance of the proposed
 PA-IHT scheme with those of the existing state-of-the-art methods for the
 TDS-OFDM system, including the modified CoSaMP based TDS-OFDM scheme \cite{Dai}
 and the DPN-OFDM based scheme \cite{DPN}. Simulation system parameters were set
 as: $f_c=643\,\rm{MHz}$, $1/T_s=7.56\,\rm{MHz}$, $N =2048$, and $M = 256$
 for the conventional TDS-OFDM transmission and $M=2\times 256$ for the DPN-OFDM
 transmission, while the perfect synchronization was assumed. We adopted the
 ITU-VB channel \cite{Chan} and the China digital television test 8th channel
 model (CDT-8) channel \cite{th} in the simulation, where both the static and
 mobile scenarios were investigated. The parameters $R_{g1}$ and $R_{g2}$ were
 adaptively set based on the channel state, while we considered $R_d=40$ in both
 the mobile and static scenarios. \begin{changebar}The simulation was carried out using
 MATLAB R2012a tool. In the simulation, the variable parameter $a$ in (13) was
 approximately chosen as $a\approx 0.1\max_{\tau_1\in D_0}\tau_1$, while the positive
 number $b$ in determining the channel sparsity level $S$ was empirically set to
 $b\in [0, ~ 5]$ where the chosen value of $b$ was inversely related to SNR.\end{changebar}

 Fig.~\ref{fig:add} shows the signal recovery probabilities as the function of the
 IBI-free region size $G$ achieved by the four different algorithms for the static
 ITU-VB channel, given $\mbox{SNR}=20$\,dB. In this simulation, if the mean square
 error (MSE) of the signal estimation was lower than $10^{-2}$, the recovery result
 was considered to be correct \cite{Dai} and hence the signal recovery probability
 was assumed to be 1. It can be clearly seen from Fig.~\ref{fig:add} that the
 proposed PA-IHT algorithm outperforms the other three algorithms significantly. The
 original IHT algorithm fails to work in this case, because its convergence requires
 that $\left\| \mathbf{\Phi} \right\|_2 < 1$ \cite{IHT}, but the measurement matrix
 (\ref{equ:phi2}) does not meet this condition. Compared with the CoSaMP algorithm
 and the modified CoSaMP algorithm which require the IBI-free region of size 40 and
 30, respectively, to recovery the signal correctly with probability 1, the proposed
 PA-IHT algorithm only needs an IBI-free region of size 7. This means that the PA-IHT
 algorithm reduces the required observation samples by $82.5\%$ and $76.7\%$,
 respectively, compared with the CoSaMP algorithm and the modified CoSaMP algorithm.
 This is because the proposed algorithm benefits from the priori information acquired,
 in terms of both the locations and the values of the partial large components in the
 target signal. Therefore, our PA-IHT algorithm are particularly effective in combating
 the CIR with a very longer delay spread, while the existing CS based schemes may
 suffer from the serious performance degradation under such adverse channel conditions.

\begin{figure}[bp!]
\vspace*{-1mm}
\begin{center}
 \includegraphics[width=\columnwidth, keepaspectratio]{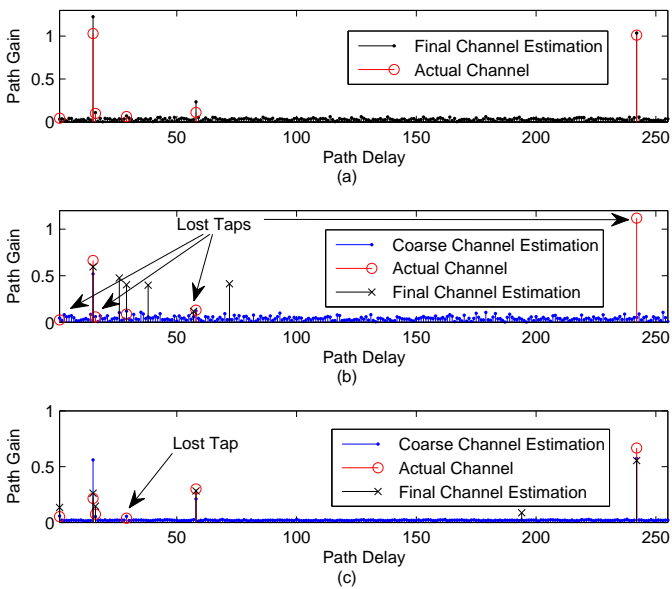}
\end{center}
\vspace*{-4mm}
 \caption{Time-domain CIR estimates of the three different schemes for the CDT-8
 channel with the mobile speed of 120km/h and given $\mbox{SNR}=10$\,dB: (a)~the
 DPN-OFDM based scheme, (b)~the modified CoSaMP based scheme, and (c)~the proposed
 PA-IHT based scheme.}
 \label{fig:ch_1} % Fig 6
\vspace*{-1mm}
\end{figure}

\begin{figure}[bp!]
\vspace*{-6mm}
\begin{center}
  \includegraphics[width=\columnwidth, keepaspectratio]{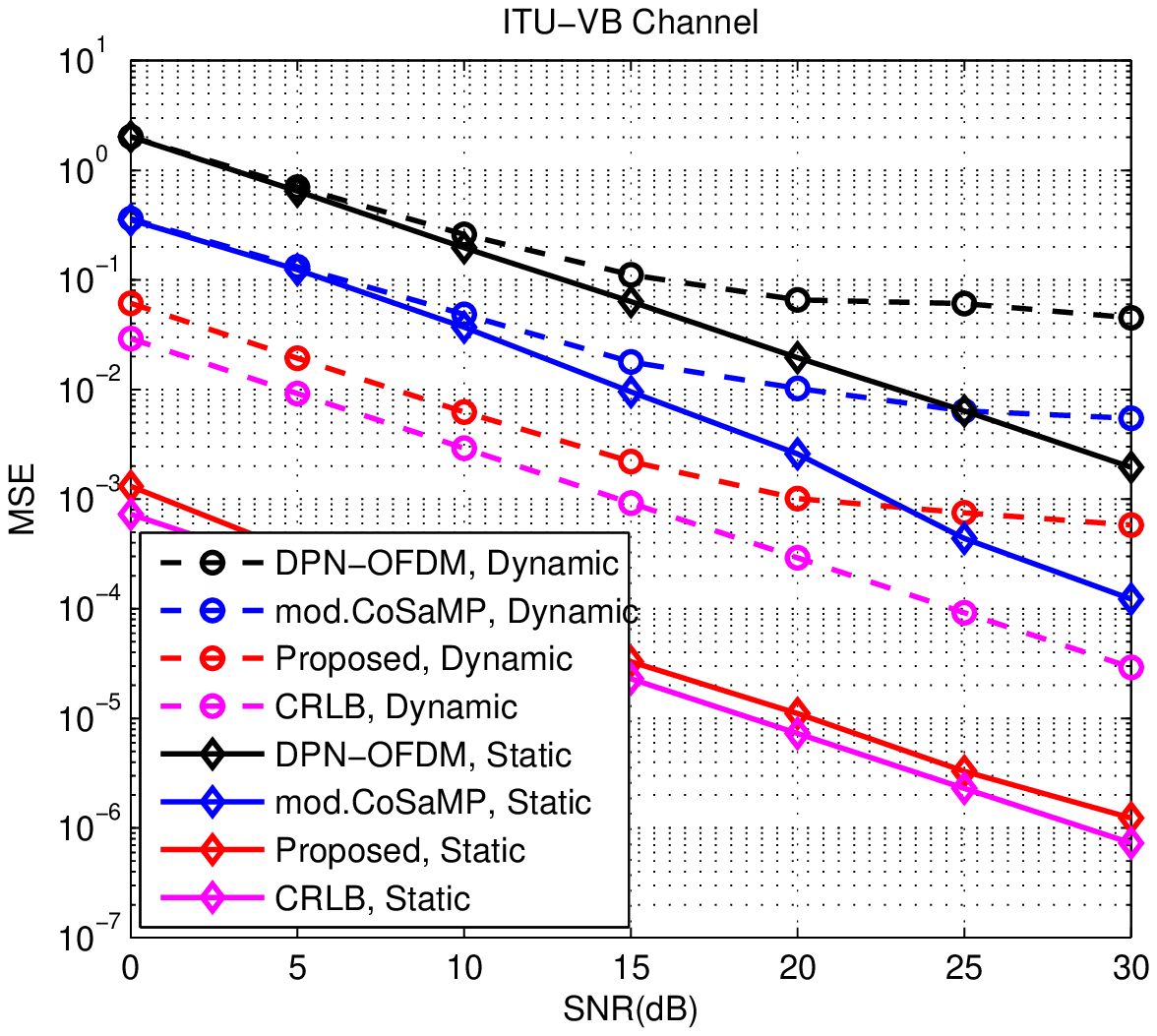}
\end{center}
\vspace*{-4mm}
\begin{center}\hspace*{5mm}{\small (a)}\end{center}
\begin{center}
  \includegraphics[width=\columnwidth, keepaspectratio]{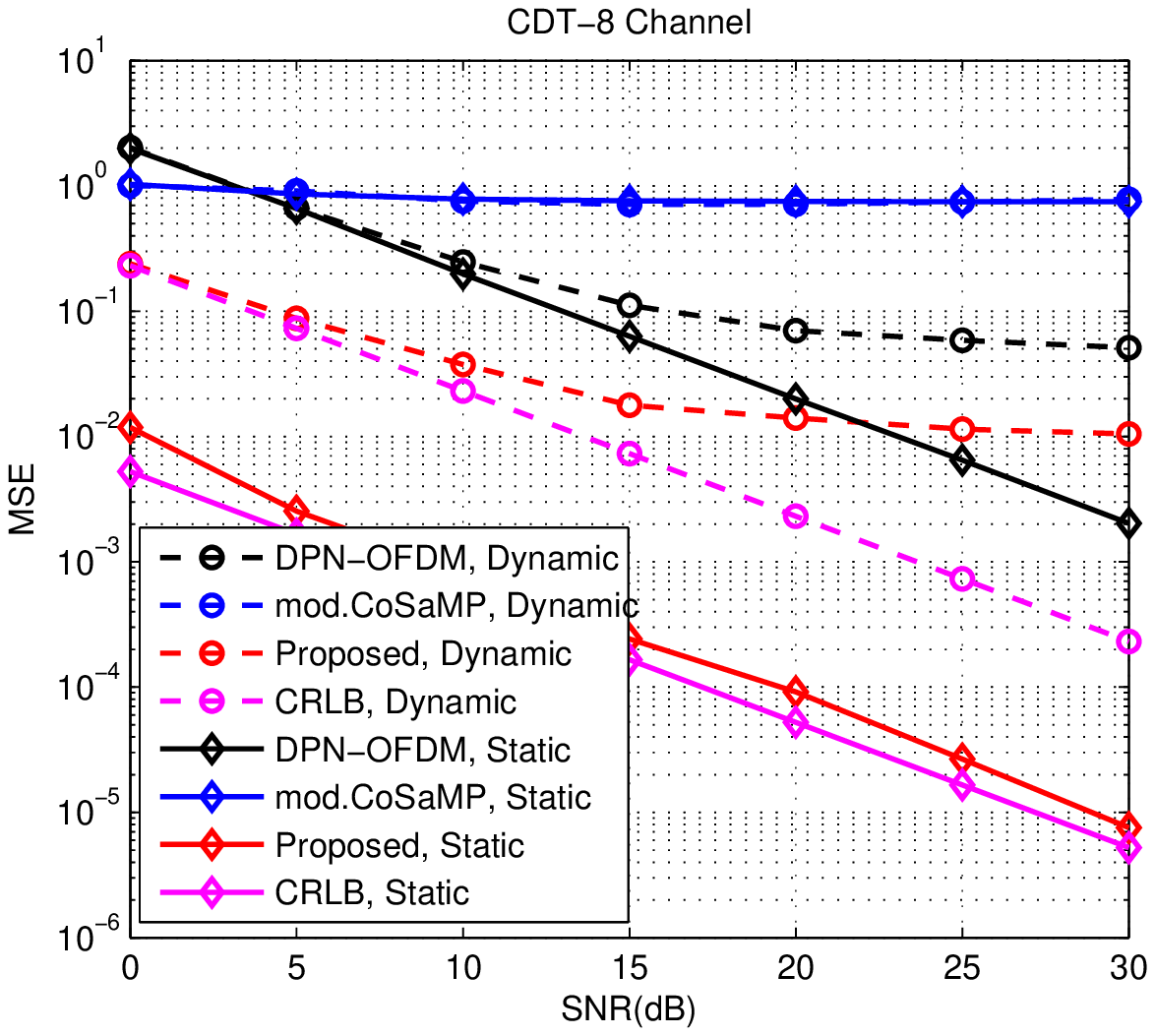}
\end{center}
\vspace*{-4mm}
\begin{center}\hspace*{5mm}{\small (b)}\end{center}
\vspace*{-4mm}
\caption{MSE performance comparison of the proposed PA-IHT scheme with the existing
 DPN-OFDM based and modified CoSaMP based schemes: (a)~the ITU-VB channel, and
 (b)~the CDT-8 channel.}
\vspace*{-1mm}
\label{fig:MSE} % Fig 7
\end{figure}

 Fig.~\ref{fig:ch_1} compares the CIR estimates obtained by the three schemes for
 the time-varying CDT-8 channel with 120km/h receiver velocity given $\mbox{SNR}=10$\,dB.
 It can be clearly seen from Fig.~\ref{fig:ch_1}~(b) that the modified CoSaMP based
 scheme performs poorly. Four actual channel path taps, including the strongest echo
 path with a long delay spread, are missing from the CIR estimate provided by this
 scheme. By contrast, only one relatively insignificant channel path tap is missing from the
 CIR estimate obtained by the proposed scheme, as can be observed from Fig.~\ref{fig:ch_1}~(c).
 This is because the CDT-8 channel has a very strong 0\,dB echo with an extremely long
 delay spread. The coarse CE method in the modified CoSaMP scheme of \cite{Dai} only
 uses the TS main part and discards the TS tail part (see the illustration of
 Fig.~\ref{fig:overlap}). Therefore, it cannot effectively detect the path delays with
 long delay spreads. By contrast, the proposed overlap-add method of the TS resolves
 this problem effectively. Moreover, by exploiting the temporal correlation of the
 wireless channel, the proposed PA-IHT scheme significantly improves the robustness
 of the coarse CE. Since the modified CoSaMP scheme \cite{Dai}  only utilises the TSs
 preceding and following the current OFDM data block, its coarse path delay detection
 may fail to work properly under instantaneous deep fading channel situations.
 From Fig. ~\ref{fig:overlap}~(a), it can be observed that the estimated gains of the
 first and fourth paths by the DPN-OFDM based scheme are lower than the noise floor.
 Hence, from the obtained CIR estimate, we cannot decide the delays of the first and
 fourth paths. Similar to the modified CoSaMP scheme, the DPN-OFDM based scheme suffers
 from serious performance degradation under instantaneous deep fading channel conditions
 as some of the estimated channel taps may be buried by the noise. Furthermore, the
 DPN-OFDM based scheme has an additional drawback of lower spectral efficiency.

\begin{changebar}
\begin{figure}[tp!]
\begin{center}
  \includegraphics[width=\columnwidth, keepaspectratio]{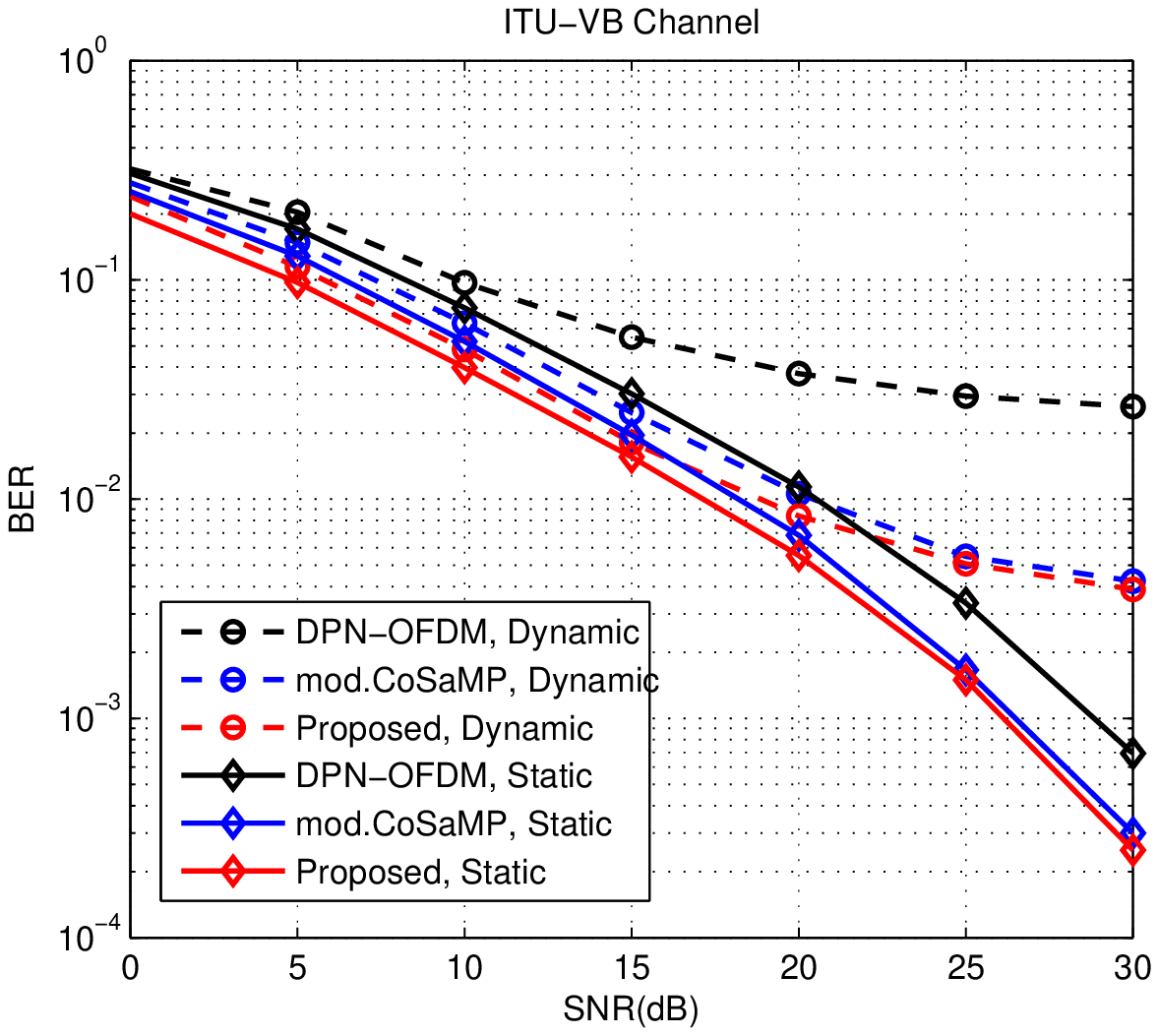}
\end{center}
\vspace*{-4mm}
\begin{center}\hspace*{5mm}{\small (a)}\end{center}
\begin{center}
  \includegraphics[width=\columnwidth, keepaspectratio]{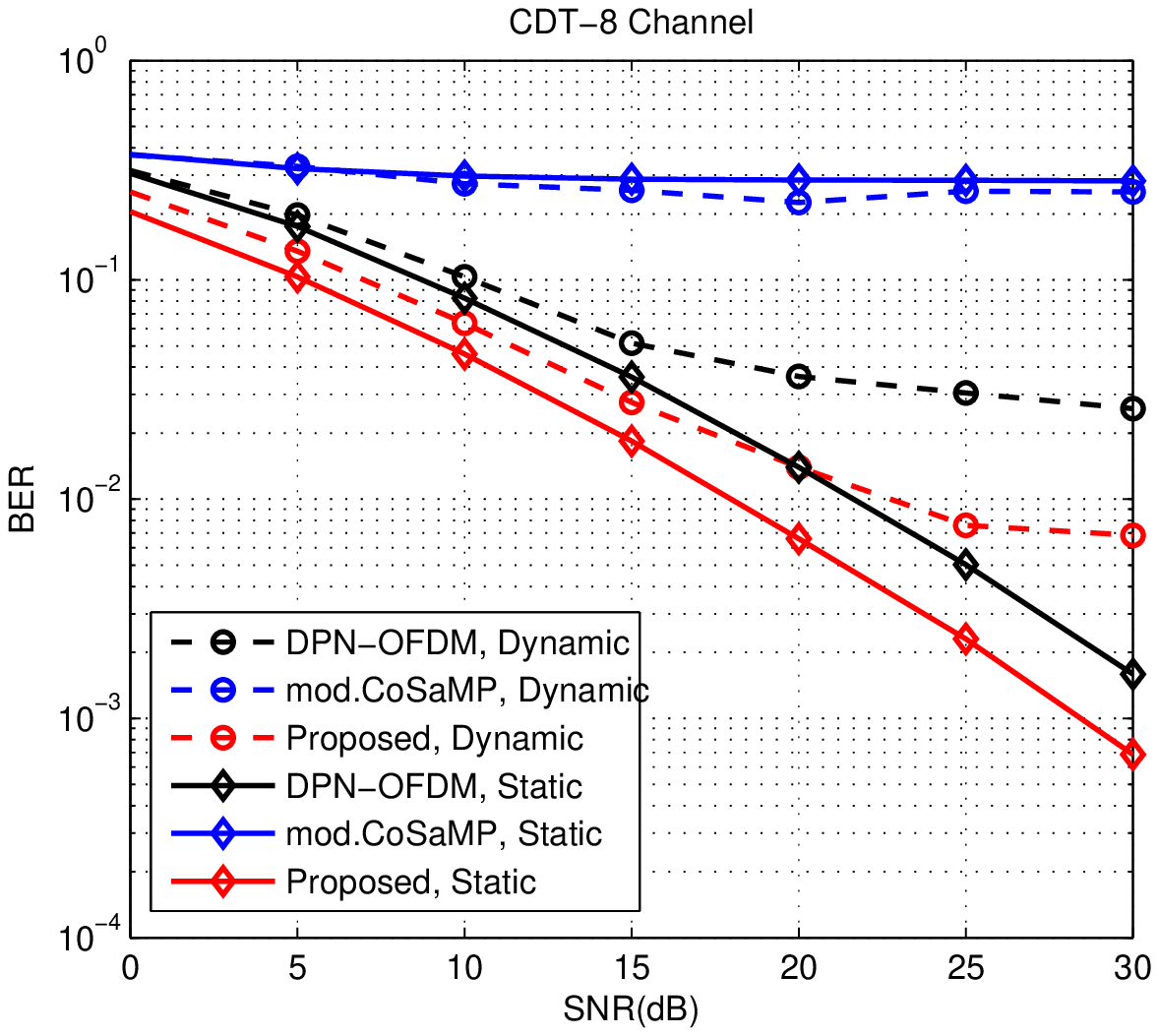}
\end{center}
\vspace*{-4mm}
\begin{center}\hspace*{5mm}{\small (b)}\end{center}
\vspace*{-4mm}
\caption{BER performance comparison of the proposed PA-IHT scheme with the existing
 DPN-OFDM based and modified CoSaMP based schemes: (a)~the ITU-VB channel, and
 (b)~the CDT-8 channel. The modulation scheme employed is QPSK.}
 \label{fig:BER} % Fig 8
\vspace*{-4mm}
\end{figure}
\end{changebar}

 Figs.~\ref{fig:MSE} and \ref{fig:BER} compare the achievable CE MSE performance
 and the data demodulation bit error rate (BER) performance of the three schemes,
 respectively, where the dynamic channel refers to the CDT-8 or the ITU-VB channel
 with the mobile speed of 120km/h. \begin{changebar}The modulation scheme employed was the
 quadrature phase shift keying (QPSK).\end{changebar} It is clear that the existing modified
 CoSaMP based scheme achieves better performance than the existing DPN-OFDM based scheme
 for the ITU-VB channel, but the modified CoSaMP based method completely fails for
 the CDT-8 channel. This again confirms that the modified CoSaMP scheme suffers
 from serious performance degradation under severe multipath propagation
 environments. The results of Figs.~\ref{fig:MSE} and \ref{fig:BER} clearly
 demonstrate that our PA-IHT based scheme significantly outperforms the two
 existing schemes in various wireless scenarios, especially in fast time-varying
 and severe multipath propagation scenarios, such as the CDT-8 channel with
 120km/h receiver velocity. More specifically, for the static ITU-VB channel, the MSE
 performance of the proposed PA-IHT based scheme are more than 20\,dB and 30\,dB better
 than the modified CoSaMP and DPN-OFDM based schemes, respectively, while it outperforms
 the other two schemes by approximately 8\,dB and 15\,dB, respectively, for the dynamic
 ITU-VB channel. For the dynamic and static CDT-8 channels, the MSE performance attained
 by our method are more than 5\,dB and 20\,dB better than the DPN-OFDM based method,
 respectively. Moreover, the MSE performance of our proposed method is very close to
 the theoretical CRLB for the two static channels. In terms of achievable BER, the
 superior performance of our scheme over the two existing ones are self-evident in
 Fig.~\ref{fig:BER}, where the performance gain of our method over the existing methods
 is particularly noticeable under doubly selective fading channel environments. This
 is owing to the following reasons. The overlap-add method of the TS based on several
 consecutive TDS-OFDM symbols significantly improves the robustness and accuracy of
 the coarse estimates for the channel length and path delays. This provides the
 accurate priori information to assist the PA-IHT algorithm. Furthermore, the sizes
 of the IBI-free region and the measurement matrix are adaptive, which further
 improves the CE accuracy of the PA-IHT algorithm. It is also worth pointing out again
 that the proposed scheme does not alter the current TDS-OFDM signal structure and it
 achieves a higher spectral efficiency than the existing DPN-OFDM scheme.

\section{Conclusions}\label{S5}

 In this paper, we have proposed a low-complexity and high-accuracy compressive
 sensing based channel estimation method, referred to as the PA-IHT algorithm,
 for widely deployed TDS-OFDM systems, which significantly outperforms the
 existing state-of-the-arts in terms of both estimation accuracy and
 computational complexity. The classical IHT algorithm for TDS-OFDM requires
 that the $l_2$ norm of the measurement matrix is smaller than 1 in order to
 guarantee convergence. By contrast, our proposed PA-IHT algorithm removes
 such a restriction and it only requires a very few iterations. We have also
 demonstrated that our scheme significantly outperforms the conventional
 DPN-OFDM based scheme. Compared with the DPN-OFDM scheme, our proposed PA-IHT
 based scheme has the additional advantage of achieving a higher spectral
 efficiency and it does not alter the current TDS-OFDM signal structure. Compared
 with the existing CS based methods, such as the modified CoSaMP algorithm for
 TDS-OFDM, our PA-IHT algorithm significantly improves the accuracy of the channel
 estimate while imposing a much lower computational complexity. Most significantly,
 our scheme maintains its effectiveness under fast time-varying severe multipath
 environments. Under such adverse channel conditions, the existing CS based scheme
 for TDS-OFDM fails to work completely.

\begin{changebar}
\section*{Appendix ~~ List of Abbreviations}
\begin{tabular}{p{13mm}p{66mm}}
 AWGN & Additive white Gaussian noise \\
 BER & Bit error rate \\
 CDT-8 & China digital television test 8th channel model \\
 CE & Channel estimation \\
 CIR & Channel impulse response \\
 CoSaMP & Compressive sampling matching pursuit \\
 CP & Cyclic prefix \\
 CRLB & Cramer-Rao lower bound \\
 CS & Compressive sensing \\
 CSI & Channel state information \\
 DFT & Discrete Fourier transform \\
 DPN & Dual pseudo-noise \\
 DTMB & Chinese digital terrestrial multimedia broadcasting standard \\
 DTTB & Digital terrestrial television broadcasting \\
 DVB-T2 & European second generation digital video broadcasting standard \\
 FFT & Fast Fourier transform \\
 IBI & Inter-block-interference \\
 IHT & Iterative hard threshold \\
 ITU-VB & International Telecommunications Union Vehicular B \\
 ML & Maximum likelihood \\
 MSE & Mean square error \\
 OFDM & Orthogonal frequency-division multiplexing \\
 PA-IHT & Priori-information aided iterative hard threshold \\
 PN & Pseudo-noise \\
 QPSK & Quadrature phase shift keying \\
 SNR & Signal to noise ratio \\
 SP & Subspace pursuit \\
 TDS & Time-domain synchronous \\
 TS & Training sequence
\end{tabular}
\end{changebar}

\end{document}